\documentclass[prb,showpacs,byrevtex,floatfix,twocolumn,superscriptaddress]{revtex4}
\usepackage{amssymb,graphicx,afterpage,amsmath,color}

\begin{document}
\title{\boldmath Rotating-crystal Malaria Diagnosis: Pre-clinical validation \unboldmath}

%
% authors and affiliations
%
%
\author{A. Orb\'an}
\affiliation{Department of Physics, Budapest University of
Technology and Economics, 1111 Budapest, Hungary}
\author{A. Butykai}
\affiliation{Department of Physics, Budapest University of
Technology and Economics, 1111 Budapest, Hungary}
\author{Zs. Pr\"ohle}
\affiliation{Department of Physics, Budapest University of
Technology and Economics, 1111 Budapest, Hungary}
\author{G. F\"ul\"op}
\affiliation{Department of Physics, Budapest University of
Technology and Economics, 1111 Budapest, Hungary}
\author{T. Zelles}
\affiliation{Department of Oral Biology, Semmelweis University,
H-1089 Budapest, Hungary}
\author{W. Forsyth}
\affiliation{Walter and Eliza Hall Institute of Medical Research,
Parkville, Victoria 3052, Australia}
\author{D. Hill}
\affiliation{Walter and Eliza Hall Institute of Medical Research,
Parkville, Victoria 3052, Australia}
\affiliation{Department of
Medical Biology, University of Melbourne, Parkville, Victoria,
Australia}
\author{I. M\"uller}
\affiliation{Walter and Eliza Hall Institute of Medical Research,
Parkville, Victoria 3052, Australia}
\author{L. Schofield}
\affiliation{Walter and Eliza Hall Institute of Medical Research,
Parkville, Victoria 3052, Australia}
\affiliation{Queensland Tropical
Health Alliance Australian Institute of Tropical Health and Medicine
James Cook University, Douglas Queensland 4811, Australia}
\author{S. Karl}
\affiliation{Infection and Immunity Division, Walter and Eliza Hall
Institute of Medical Research, Parkville, Victoria 3052, Australia}
\author{I. K\'ezsm\'arki}
\affiliation{Department of Physics, Budapest University of
Technology and Economics, 1111 Budapest, Hungary}
\affiliation{Condensed Matter Research Group of the Hungarian
Academy of Sciences, 1111 Budapest, Hungary}
\date{\today}
\begin{abstract}
Improving the efficiency of malaria diagnosis is one of the main
goals of current malaria research. We have recently developed a
magneto-optical (MO) method which allows high-sensitivity detection
of malaria pigment (hemozoin) crystals via their magnetically
induced rotation in blood. Here, we validate this technique on
laboratory derived blood samples infected with \textit{Plasmodium
falciparum}. Using two parasite cultures, the first containing
mostly ring stages and the second corresponding to the end of the
parasite life cycle, we demonstrate that our novel method can detect
parasite densities as low as $\sim$40 and $\sim$10\,parasites per
microliter of blood for ring and schizont stage parasites,
respectively. This detection limit exceeds the performance of rapid
diagnostic tests and competes with the threshold achievable by light
microscopic observation of blood smears. Our method can be performed
with as little as 50\,microliter of capillary blood and is sensitive
to the presence of hemozoin micro-crystals down to ppm
concentrations. The device, designed to a portable format for
clinical and in-field tests, requires no special training of the
operator or specific reagents, except for an inexpensive lysis
solution to release intracellular hemozoin. Beyond diagnostics, this
technique may offer an efficient tool to study hemozoin formation,
trace hemozoin kinetics in the body and test
susceptibility/resistance of parasites to new antimalarial drugs
inhibiting hemozoin formation.
\end{abstract}
\maketitle

\section*{Introduction}
Although there is a plethora of emerging techniques aiming at
high-sensitivity diagnosis of malaria, only a few of these
approaches are feasible for clinical and in-field diagnosis. Apart
from purely symptom based, presumptive diagnosis, the two main
diagnostic methods currently in practice are the antigen-based
detection of malaria parasites using rapid diagnostic tests (RDT)
and the microscopic observation of infected red blood cells in blood
smears.\cite{Moody2002,Hanscheid1999,Payne1988,Wongsrichanalai2007}
The detection limits of RDT and light microscopy  have been reported
to be approximately 100\,parasites/$\mu$L and
5-50\,parasites/$\mu$L,
respectively.\cite{Moody2002,Maltha2013,Alonso2011,Prudhomme2006}
Both of these methods are subject to inherent limitations: i)
although RDTs are becoming more affordable, they cannot provide a
quantitative measure of parasitemia and presently do not possess
sufficient sensitivity to detect low-level infections which are very
common in endemic settings, ii) the visual inspection of blood
smears is time and labor intensive. Moreover, the detection
threshold of 5\,parasites/$\mu$L is rather theoretical and can only
be achieved under ideal conditions (good-quality blood film, highly
trained microscopist, high-powered microscope, etc.). In practice,
most routine diagnostic laboratories achieve approximately
50\,parasites/$\mu$L and detect about 50\,\% of malaria
cases.\cite{Alonso2011,Okell2009,Perkins2008}

Among molecular biology-based methods, polymerase chain reaction
(PCR) assays surpass the performance of RTDs and light
microscopy.\cite{Coleman2006,Snounou1996} However, they often
require expensive equipment and reagents, highly trained laboratory
personnel and are prone to contamination.\cite{Tangpukdee2009}
Recent studies conclude that real-time PCR has a detection limit
corresponding to a few parasites in 1\,$\mu$L
blood,\cite{Owusu2013,Khairnar2009} nevertheless, it is not yet a
practical method for routine diagnosis under field conditions.

The idea to take advantage of the unique magnetic properties of
malaria pigment (hemozoin) and to use it as an alternative target of
optical diagnosis has been proposed by several
groups.\cite{Butykai2013,Karl2008,Mens2010,Newman2008,Zimmerman2006}
Hemozoin is a micro-crystalline heme compound produced by malaria
parasites as they detoxify free heme derived from hemoglobin
digestion. Our recent study using synthetic hemozoin crystals
suspended in blood demonstrated that the rotating-crystal
magneto-optical (MO) diagnostic method can detect hemozoin
concentrations down to 15\,pg/$\mu$L.\cite{Butykai2013} This
threshold concentration was estimated to be equivalent to a parasite
density of $\leq$30\,parasites/$\mu$L in infected blood provided
that the whole amount of hemozoin produced by the parasites is
released into the lysed cell suspension.

However, the relation between hemozoin concentration and parasite
density in human infections is not straightforward.
Intraerythrocytic hemozoin content is dependent on the maturity of
the blood stage parasites; with the least amount of hemozoin present
in erythrocytes during the ring stage and the highest amount during
the schizont stage.\cite{Moore2006,Orjih1993,Becker2004} Therefore,
the hemozoin concentration in blood derived from human infections
depends on the parasite stage distribution at the time the blood
sample was collected. In \textit{P. falciparum} (the most lethal
parasite species) infections often only the ring and early
trophozoite stages are found in the peripheral circulation, since
the later developmental stages cytoadhere to the vascular
endothelium.\cite{Cooke1995}

Moreover, the MO signal recorded by our technique depends not only
on the amount but also on the size and morphology of the hemozoin
crystals, which can be different for synthetic and naturally grown
crystals.\cite{Slater1991,Noland2003,Jaramillo2009} The MO method is
sensitive only to those hemozoin crystals released into suspension,
which can be magnetically rotated. Correspondingly, aggregation of
crystals or their binding to other components of lysed infected
blood, such as cell membranes, could substantially decrease the
sensitivity of our technique. Signal loss may be avoided by
appropriate lysis and treatment of blood samples prior to
measurement.

In the present study we aimed to address these key issues and to
validate the rotating-crystal malaria diagnosis method using
synchronized cultures of \textit{P. falciparum}. (For a short
description of the detection scheme see Materials and Methods
section.) For this purpose, we investigated its sensitivity and
detection threshold for two cultures with different maturity
distributions of the parasites. The first, hereafter referred to as
the \textit{ring stage culture}, contained mostly ring stages and
some early trophozoites with a total parasite density of
P$\approx$3.1$\times$10$^5$\,parasites/$\mu$L. It is representative
to the distribution of parasite blood stages most often encountered
in \textit{P. falciparum} infections. The second with a lower total
parasite density of P$\approx$2.8$\times$10$^4$\,parasites/$\mu$L
corresponds to the end of the parasite life cycle where some of the
parasites are still in the schizont form but most of them, following
an invasion, already turned to early-stage rings of the next
generation. The distributions of the parasites among the different
stages -- early-ring, late-ring, early-trophozoite,
late-trophozoite, early-schizont and late-schizont stages -- in the
two cultures are displayed in Fig.~1 together with light microscope
images of parasites representative to these stages. Since the
early-stage rings do not contain hemozoin, the hemozoin content
present in the second culture is formed in the first cycle before
the invasion mostly during the schizont
stage.\cite{Moore2006,Orjih1993,Becker2004} For this reason,
hereafter we refer to this culture as the \textit{schizont stage
culture}. Schizont stage parasites are not normally found in the
peripheral blood during human \textit{P. falciparum} infection due
to their sequestration in small capillary blood
vessels.\cite{Cooke1995} However, this stage restriction is not
present in other non-sequestering species of human malaria parasites
such as \textit{P. vivax}.\cite{Carvalho2010}

\begin{figure}[!ht]
\includegraphics[width=2.7in]{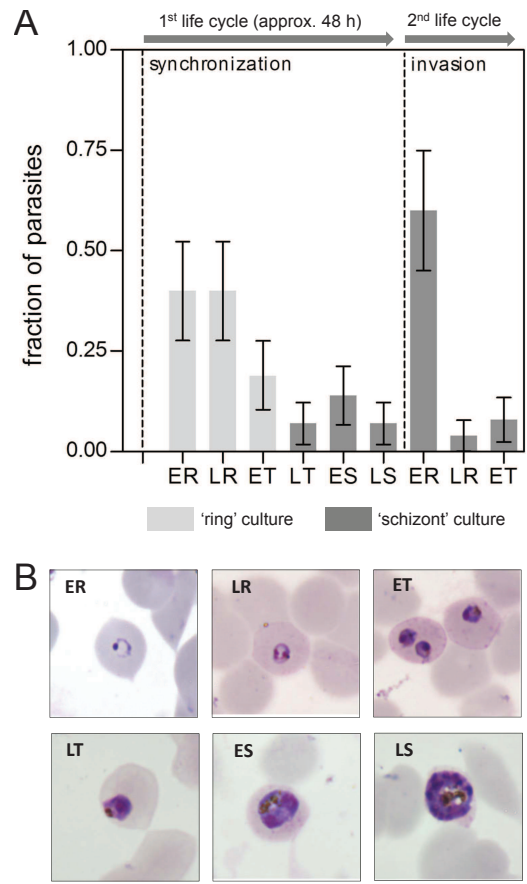}
\caption{\textbf{Distribution of parasite life cycle stages in the
two \textit{Plasmodium falciparum} cultures used in the present
study.} Panel A: The \textit{ring stage culture} contained early
rings, late rings and some early trophozoites of the first
generation after synchronization, while the \textit{schizont stage
culture} corresponded to the end of the first life cycle where most
of the schizont stages have already turned to early ring stages of
the second generation after synchronization. Therefore, the ring
stage culture contained only the hemozoin present in the parasites
up to the early trophozoite stage, while the schizont stage culture
had the entire hemozoin content formed over one generation of
parasites with the largest portion produced  by schizonts. Panel B:
Light microscopy images of Giemsa stained thin blood films
containing infected red blood cells with parasites in different
stages of maturity (taken from these two cultures). In both panels
the labels ER, LR, ET, LT, ES and LS correspond to early-ring,
late-ring, early-trophozoite, late-trophozoite, early-schizont and
late-schizont stages, respectively.}
\end{figure}
\begin{figure*}[!t]
\includegraphics[width=7in]{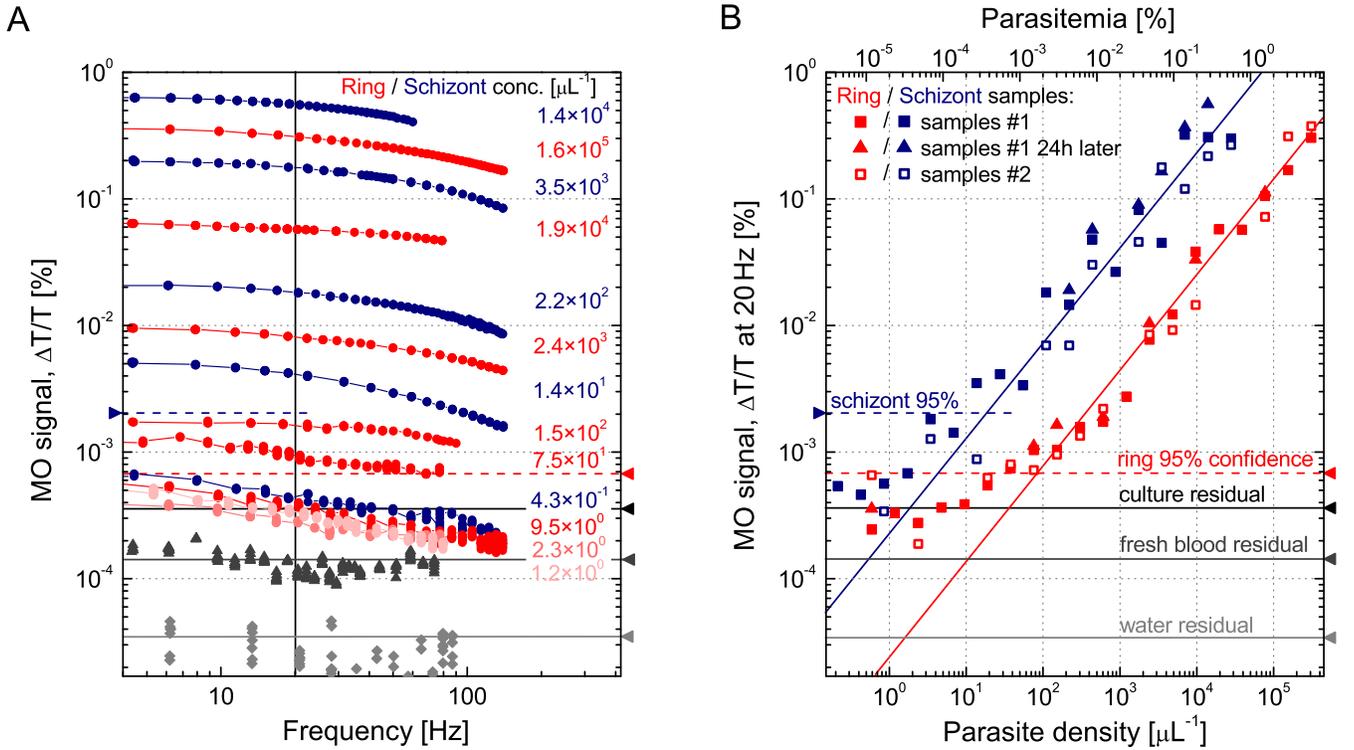}
\caption{\textbf{Magneto-optical (MO) detection of parasitemia in
synchronized \textit{Plasmodium falciparum} cultures.} Panel A: Red
and blue curves show the frequency dependent MO signal for samples
from the ring and schizont stage cultures, respectively, with
various levels of parasite density given in $\mu$L$^{-1}$ units on
the right of the respective curves. For ring stage samples with
parasite densities lower than 10\,parasites/$\mu$L the MO signal
does not further decrease. Data plotted with triangles and diamonds
are the residual signal from freshly hemolyzed uninfected blood and
water, respectively. The frequency scale denotes twice the rotation
frequency of the magnetic field, i.e. the frequency of the modulated
intensity. Panel B: Red and blue squares in panel B are the MO
signal values measured at 20\,Hz -- indicated by a vertical solid
line in panel A -- for the dilution series prepared from the
original ring and schizont stage cultures, respectively. Solid and
open squares correspond to the duplicate samples labeled as samples
\#1 and samples \#2. Triangles indicate the results obtained by
remeasuring samples \#1 with 24\,h delay. The solid lines following
the trend of the MO signal at higher parasite densities for ring
(red line) and schizont (blue line) samples are guides for the eye.
The black horizontal line represents the mean residual MO signal for
the cultures, while the 95\,\% confidence levels of this mean
detection limit for the ring and schizont stage samples are
indicated by red and blue dashed lines, respectively.
Correspondingly, for ring and schizont stage samples with parasite
density higher than 40\,parasites/$\mu$L and 10\,parasites/$\mu$L,
respectively, the diagnosis is positive with a confidence of at
least 95\,\%. The background signal for freshly hemolyzed uninfected
blood and water are also shown by dark and light grey lines. All
these horizontal indicators are also shown in panel A for reference.
The upper horizontal scale shows the corresponding levels of
parasitemia.}
\end{figure*}

\section*{Results}
The MO signal, the measure of hemozoin content within the lysed cell
suspension, is shown in Fig.~2 for dilution series of the ring and
schizont stage cultures. The 20 serial 2-fold dilutions using
uninfected erythrocytes allowed for MO signal to be assessed over 6
orders of magnitude of parasitemia. As a general trend in Fig.~2A,
the MO signal varies proportionally to the parasitemia level and the
signal for each sample shows a gradual decrease with increasing
frequencies of the rotating magnetic field. This frequency
dependence is in agreement with previous results obtained when using
synthetic hemozoin crystals suspended in blood and originates from
the viscosity of the lysed cell suspension hindering fast rotations
of the crystals.

As schizont stages contain more hemzoin than ring stage-parasites,
samples from the diluted schizont stage culture exhibited higher
signals than ring stage samples with comparable parasitemia.
Although the overall frequency dependence is similar for the two
cultures, the decrease in the signal with increasing frequency is
more pronounced for schizont stage samples, which is likely due to
the larger crystal size in these samples. At low levels of parasite
density, namely for samples with $P$$\leq$10\,parasites/$\mu$L from
the ring stage culture, the signal does not further drop with
decreasing parasitemia. The frequency dependence of the MO signal
for these ring stage samples also becomes different; the
low-frequency saturation common for higher concentrations does not
hold anymore. These imply a residual MO signal not related to
hemozoin.

In order to determine the detection limit of our method, the results
obtained for the dilution series of the ring and schizont stage
cultures are summarized in Fig.~2B, where the MO signal at 20\,Hz is
plotted versus the parasitemia (and parasite density). Sequential
measurements performed on the same sample with time delays less than
one hour gave identical results. In several cases we checked the
reproducibility of the protocol by repeating the measurement for
both of the duplicate samples labeled as samples \#1 and samples \#2
in Fig.~2B. For ring stage samples with
$P$$\leq$10\,parasites/$\mu$L, where the MO signal shows no
systematic variation with decreasing parasite density, we calculated
the mean value of the residual MO signal
($\approx$3.6$\times$10$^{-4}$\,\%) and its standard deviation
($\approx$1.5$\times$10$^{-4}$\,\%). Assuming Gaussian distribution
for the residual MO signal values, we found that the 95\,\%
confidence level of the mean detection limit for ring stage samples
is $\Delta T/T$$\approx$6.6$\times$10$^{-4}$\,\%, which corresponds
to a parasite density of $\sim$40\,parasites/$\mu$L. This is
equivalent to the parasitemia level of 8$\times$$10^{-4}$\,\%. Since
the reproducibility is poorer for schizont stage samples, in this
case our rough estimate for the 95\,\% confidence level of the mean
detection limit is considerably higher with $\Delta
T/T$$\approx$2$\times$10$^{-3}$\,\%, which corresponds to the
parasite density of approximately $\sim$10\,parasites/$\mu$L and a
parasitemia level of $\sim$2$\times$$10^{-4}$\,\%. Note that the
reproducibility between duplicate samples does not significantly
vary with parasite density for dilutions of either the ring or the
schizont stage cultures.

The hemozoin content of the two cultures can be roughly estimated
from the parasite density and the stages of parasite development
specified in Fig.~1. Ring and early trophozoite stages up to 24\,h
were reported to convert about 3-15\,\% of the total hemoglobin in
the infected red blood cells to hemozoin,\cite{Moore2006,Orjih1993}
while at the schizont stage this portion is increased to
approximately 50-80\,\%.\cite{Francis1997,Hackett2009,Weissbuch2008}
According to the 3-15\,\% hemoglobin conversion rate reported for
rings and early trophozoites, the undiluted ring stage culture with
total parasite density of P=3.1$\times$10$^5$\,parasites/$\mu$L
contained $\sim$9-48\,ng/$\mu$L hemozoin. The undiluted schizont
stage culture, which had about 10 times lower parasite density with
P=2.8$\times$10$^4$\,parasites/$\mu$L, contained all the hemozoin
formed during the first cycle. Using the hemoglobin conversion rates
quoted above, this corresponds to $\sim$14-23\,ng/$\mu$L hemozoin.
An independent estimate, based on the MO signal yields approximately
6\,ng/$\mu$L and 9\,ng/$\mu$L for the undiluted ring and schizont
stage cultures, respectively. For this estimate, we used the
conversion factor c$_{Hz}$=1\,ng/$\mu$L $\rightarrow$ $\Delta
T/T$=1.4\,\% between the hemozoin concentration and the
low-frequency ($\sim$1\,Hz) MO signal previously determined for
artificial hemozoin crystals suspended in blood.\cite{Butykai2013}
Note that the 20-fold dilution of the samples prior to the MO
measurement needs to be taken into account, since this conversion
factor applies for samples with 50\,\% hematocrit.

We also estimate the hemozoin concentration of the two undiluted
cultures based on MO signal using the conversion factor
c$_{Hz}$=1\,ng/$\mu$L $\rightarrow$ $\Delta T/T$=1.4\,\% between the
hemozoin concentration and the low-frequency ($\sim$1\,Hz) MO signal
previously determined for artificial hemozoin crystals suspended in
blood \cite{Butykai2013}. This yields approximately 6\,ng/$\mu$L and
9\,ng/$\mu$L for the undiluted ring and schizont stage cultures,
respectively. Note that the 20-fold dilution of the samples prior to
the MO measurement needs to be taken into account, since this
conversion factor applies for samples with 50\,\% hematocrit.

\begin{figure}[!t]
\includegraphics[width=3.4in]{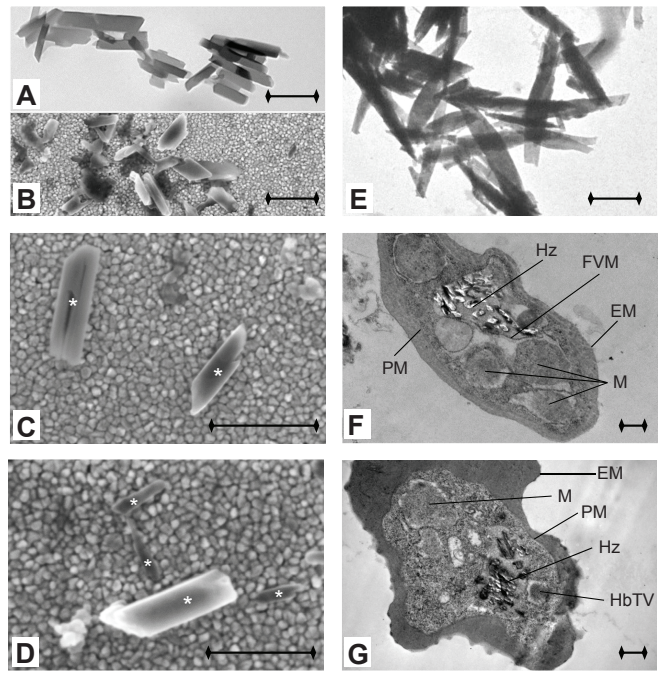}
\caption{\textbf{Hemozoin crystals observed by electron microscopy
in-situ and after isolation.} The transmission electron microscopy
(TEM) image in panel A and scanning electron microscopy (SEM) images
in panels B, C and D show hemozoin crystals extracted from the
samples previously used for the MO measurements. The granular
background of the SEM images comes from the gold coating of the
glass substrate. Individual crystals in panels C and D are marked
with asterisks. The length of the black bar at the bottom right
corner of each panel is 500\,nm. The typical length of the extracted
hemozoin crystals ranges from 200\,nm to 500\,nm. For comparison, a
TEM image of synthetic hemozoin crystals used in our previous MO
study\cite{Butykai2013} is shown in panel E. These synthetic
crystals have more elongated shape with a typical length of
500-900\,nm. Panels F and G display TEM images of intact infected
erythrocytes in the schizont stage. Distinct components of the
parasite and the erythrocyte can be observed including the
erythrocyte membrane (EM), parasite membrane (PM), food vacuole
membrane (FVM), merozoites (M), hemoglobin transport vesicles
(HbTV), knobs (K) and hemozoin (Hz).}
\end{figure}

The first scenario is supported by the electron microscopy images in
Fig.~3, where hemozoin crystals inside of infected erythrocytes and
as extracted from the cultures are shown. These elongated
crystallites are considerably smaller (with typical length of
$\sim$200-500\,nm) than the synthetic ones ($\sim$500-900\,nm)
previously studied and also displayed in the
figure.\cite{Butykai2013} This is also in accordance with the weaker
frequency dependence of the MO signal observed in the present study,
which indicates that natural crystals are able to follow the
rotation of the magnetic field up to higher frequencies than the
synthetic ones as a possible consequence of their reduced size.
Furthermore, the comparison between the natural crystals within the
parasites and those extracted from the cultures confirms no major
change either in the size or in the morphology of the crystals due
to our lysis-sonication protocol.

In order to minimize binding and aggregation of crystals we repeated
the measurements 24\,h later following another 30\,min of
sonication. We found modest but systematic increase of the MO signal
of typically not more than 10-30\,\%. On this basis, we expect that
most of the hemozoin is successfully released into suspension,
likely in the form of individual crystals.

The presence of a frequency-dependent residual MO signal indicates
that some components of lysed blood can be magnetically oriented and
rotated similarly to the hemozoin crystals. We have also studied
freshly drawn blood following the same lysis protocol. The residual
MO signal observed in this case was considerably lower than found
for either the ring or schizont stage samples (see Fig.~2), which
implies better detection threshold for instant diagnosis. We suspect
that due to the freeze-thaw-lysis procedure applied to the present
set of samples, some portion of the hemoglobin may have been
transformed to an aggregated or polymerized form -- similar to the
intracellular non-covalent polymerization of hemoglobin previously
observed in sickle cell
disease\cite{Noguchi1983,Noguchi1984,Ofori2001} --, which may
produce the residual MO signal. These points, requiring additional
systematic studies, stress the crucial role of an appropriate blood
treatment prior to diagnosis.

The noise floor of our equipment, determined using pure water, is
roughly frequency independent and about one order of magnitude
smaller than the residual signal from fresh blood. This enables
further improvement of the detection limit provided that the
residual MO signal from blood can be reduced by optimizing blood
treatment. Optimizing the properties of the lysis solution may also
help to dissociate the crystals without the need for sonication.

\section*{Discussion}
The potential of exploiting hemozoin as a magnetic biomarker for
malaria diagnosis has stimulated extended research over the last few
decades. Taking advantage of the paramagnetic nature of hemozoin,
several approaches have been proposed to improve the sensitivity of
existing methods by the magnetic separation of malaria infected
erythrocytes from whole blood prior to the
diagnosis.\cite{Paul1981,Carter2003,Karl2008,Karl2011} More
recently, new techniques have been emerging, which directly use
hemozoin as a target material of magnetic diagnosis. These
techniques include detection of depolarized side-scatter in flow
cytometry,\cite{Frita2011} electrochemical magneto
immunoassays,\cite{Castilho2011} magnetically enriched surface
enhanced resonance Raman spectroscopy\cite{Yuen2012,Hobro2013} and
magneto-optical detection using polarized
light.\cite{Mens2010,Newman2008,Newman2010} Among them, to the best
of our knowledge, our rotating crystal MO diagnostic device is the
first realized in a cost-effective portable format with excellent
sensitivity.

In the present study, the detection limit of our rotating-crystal MO
diagnostic device was found to be $\sim$40\,parasites/$\mu$L and
$\sim$10\,parasites/$\mu$L for ring and schizont stage parasites,
respectively. These detection limits are below the threshold
currently achievable with RDTs ($>$100\,parasites/$\mu$L) and lie
within the same range as the limits of conventional optical
microscopy for malaria diagnosis
(5-50\,parasites/$\mu$L).\cite{Moody2002} For the present set of
blood samples kept frozen and thawed before the measurement, the
performance of the method was limited by a residual MO signal due to
some part of the lysed cell suspension. This residual MO signal
obscures the genuine MO signal of hemozoin at parasite densities
lower than the limits quoted above. Preliminary results indicate
that for measurements on freshly lysed blood samples, which is the
condition relevant to instant diagnosis, the detection limit of our
rotating crystal MO platform could be further improved. Thus, this
methodology has the potential to yield portable tools for instant
diagnosis with detection limits approaching that of PCR-based
platforms.

Limitations of our diagnostic technique include i) the possibility
of false positive detections due to the presence of hemozoin in the
blood, e.g. contained within white blood cells,\cite{Schwarzer2001}
for extended periods of time after an infection has been cleared and
ii) the possibility of false negative results in case an infection
only contains very early ring stage parasites with little or no
hemozoin. Furthermore, methods targeting only hemozoin as a marker
for infection are thought to have limitations in their diagnostic
capacity as they cannot distinguish between different malaria
species. The specificity of our MO diagnostic scheme, owing to
variations in the typical size and morphology of hemozoin crystals
produced by different species, needs to be tested by a comparative
study on various \textit{Plasmodium strains}. We emphasize that only
studies on field isolates will be able to elucidate the impact of
these possible confounding factors and the present study is the
basis for such field-based trials.

It is currently believed that without active case detection of
asymptomatic malaria infections, malaria eradication will be
impossible or very difficult to achieve.\cite{Alonso2011} However,
there are no diagnostic tools for rapidly screening hundreds of
people per day, on-site and with high sensitivity.\cite{Baird2010}
The rotating-crystal MO diagnostic device has the potential to
fulfill these requirements as it is cost-effective, rapid, highly
sensitive, portable and easy to apply.

Besides on-site diagnosis, the present methodology provides an
efficient in-vitro laboratory tool to test the susceptibility of the
parasites to new antimalarial drugs by monitoring the effect of
treatment on the rate of hemozoin formation. The scope of the
technique described here may also cover the study or diagnosis of
other human diseases, such as schistosomiasis, which are also caused
by blood-feeding organisms producing hemozoin similarly to malaria
parasites.\cite{Chen2001,Oliveira2004,Oliveira2000,Karl2013}\\
\vspace{0.3in}

\section*{Materials and Methods}
\subsection*{Parasite culture} \textit{P. falciparum} parasites
(laboratory adapted strain 3D7) were cultured following the method
of Trager and Jensen with modifications \cite{Trager1976}. The
culture medium was RPMI 1640 with L-glutamine (GIBCO cat \# 31800)
supplemented with 2\,mg/mL NaHCO$_3$ (Merck, cat \# 106329),
25\,mg/L gentamicin (Pfizer, cat \# 61022027), 50mg/L hypoxanthine
(Calbiochem, cat \# 4010), 25\,mM HEPES (SAFC, cat \# 90909C) and
10\,\% pooled O+ human serum (mixed blood groups, Australian Red
Cross Blood Service). Cultures were maintained at 4\,\% hematocrit
with changes of culture medium every 48\,h and diluted with
uninfected O+ red blood cells when the parasitemia exceeded 5\,\%.
Parasites were maintained in an atmosphere of 5\,\% CO$_2$ and 1\,\%
O$_2$ in N$_2$. Parasite cultures were kept in stage synchrony by
applying the 5\,\% Sorbitol method, first described by Lambros and
Vanderberg \cite{Lambros1979}. Hemozoin liberated from late stage
parasites by the synchronization process was removed  by washing the
cells in RPMI medium after the Sorbitol induced cell lysis and
before re-establishment of the culture.

Parasite densities for the the ring and schizont stage cultures
described above were estimated by counting the number of parasites
contained in 5000 red blood cells on Giemsa stained thin blood
films. The commonly used conversion from parasitemia to parasite
density based on the assumption that 1\,$\mu$L of blood contains
5$\times$$10^6$ red blood cells at 50\% hematocrit was applied
\cite{Moody2002}. For both, the ring and schizont stage cultures,
2-fold dilution series were prepared in duplicate in uninfected
human  erythrocytes (Red Cross blood bank, Royal Melbourne Hospital,
VIC, Australia). The duplicate dilution series prepared for ring and
schizont cultures contained 21 and 20 dilutions, respectively, where
each of the dilutions had a volume of 200\,$\mu$L at a hematocrit of
50\,\%. These dilutions were immediately frozen and thawed twice to
create lysates. These lysates, hereafter referred to as ring and
schizont stage samples, were subsequently frozen at -80$^o$C and
kept in a frozen state until they were thawed immediately prior to
measurement. The personnel carrying out the MO measurements were
blinded against the contents of each of the samples to reduce
potential observer bias.

\subsection*{Blood treatment prior MO diagnosis} Blood samples
prepared for MO measurement were thawed at room temperature and were
diluted 20-fold with distilled water, resulting in a total volume of
4\,mL per specimen. Additionally, 100\,$\mu$L of a special red cell
lysis buffer, hereafter referred to as clearing solution, was added
to the specimens in order to disperse the remnants of the lysed
RBCs. In preliminary experiments, this clearing solution (2.5\,V/V\%
Triton X-100 in 0.1\,M NaOH) was confirmed not to cause noticeable
degradation of hemozoin when added to synthetic malaria pigment
suspended in lysed blood. Note that the final concentration of NaOH
in the samples was only 2.5\,mM. The samples were subjected to
30\,min of ultrasonication to dissociate potential aggregates of
hemozoin and ensure an unhindered motion of crystals in the fluid.
Owing to this treatment a transmittance of $\sim$30-40\,\% could be
achieved with no substantial light scattering observed from the
blood samples. MO signals were recorded on 1\,mL volumes taken from
the samples subsequent to the preparation process and
reproducibility was confirmed in several cases after one day of
storage of the samples at 4$^o$C and by carrying out the measurement
on both of the duplicate samples.

\subsection*{Magneto-optical measurements} MO measurements were
performed with the prototype of the rotating magnet setup described
in our previous study \cite{Butykai2013}. We utilize a permanent
magnetic ring, which produces a B=1\,T magnetic field at the sample
position and can be rotated with adjustable frequency. Polarized
light from a laser diode is transmitted through the sample in the
direction perpendicular to the plane of the rotating magnetic field.
Owing to the magnetic alignment of the freely rotating  and dichroic
hemozoin crystals present in infected blood, magnetically induced
linear dichroism can be observed and quantified as the difference in
transmission for light polarized along and perpendicular to the
magnetic field direction ($\Delta T$) divided by the average
transmission ($T$), i.e. $\Delta T/T$ \cite{Butykai2013}. Due to the
synchronous periodic rotation of the crystals the linear dichroism
gives rise to a periodic change in the transmitted intensity
oscillating with twice the magnet rotation frequency. This second
harmonic intensity component ($I_{2f}$) can be effectively filtered
by a lock-in technique. The MO signal is then obtained as the ratio
of the modulated light intensity and the average intensity
($I_{dc}$); $\Delta T/T=I_{2f}/4I_{dc}$.

This detection scheme for malaria diagnosis provides the best
signal-to-noise ratio in the rotation frequency regime of 10-30\,Hz.
In order to exclude baseline artifacts emerging from mechanical
vibration or improper optical alignment of the system, the baseline
is checked with pure distilled water prior to the measurement of
each sample. This water baseline -- due to electronic and mechanical
noise -- was generally found to be almost two orders of magnitudes
lower than the signal from the samples with the lowest hemozoin
concentration.

\subsection*{Electron microscopy} For transmission electron
microscopy (TEM), parasite samples were fixed in resin blocks and
70-120\,nm thin sections were cut using a Leica EM UC6 microtome
(Leica Microsystems, North Ryde, NSW, Australia) and brought onto
carbon coated copper TEM grids (ProSciTech, Thuringowa, Qld.,
Australia). The TEM grids were then stained with 5\,\% uranyl
acetate for 15\,min and Reynold's lead citrate solution for 5\,min.
TEM was conducted on a JEOL 2100 TEM (JEOL Inc., Tokyo, Japan). For
details of the method see the Supporting Information.

For scanning electron microscopy (SEM), the samples giving the
highest MO signal were used and hemozoin crystals were extracted
following the method of Chen and coworkers \cite{Chen2001}. The dark
brown pellet obtained by this method was resuspended in 80\,$\mu$L
water. For SEM imaging small droplets of the suspension containing
the hemozoin crystals were applied to gold coated glass slides
without further purification or treatment. The droplets were dried
overnight at room temperature. The SEM images were acquired on a LEO
1540XB electron microscope using the in-lens detector. The
accelerating voltage was set to 3\,kV and the viewing angle was
perpendicular to the gold surface.

\textbf{Acknowledgements.} The authors acknowledge fruitful
discussions with T. Hanscheid and assistance with electron
microscopy by L. Kyriliak, M. Saunders, J. Shaw and facilities of
the Centre of Microscopy, Characterization and Analysis at The
University of Western Australia. This work was supported by
Hungarian Research Funds OTKA K108918,
T\'AMOP-4.2.1.B-09/1/KMR-2010-0001, by NHMRC grants GNT1021544 and
GNT1043345 awarded to IM and by NHMRC grants GNT637406 and
GNT1058665 awarded to LS. SK is supported through a NHMRC early
career research fellowship (GNT1052760).\\
\vspace{0.1in}

\textbf{Correspondence:} Istvan Kezsmarki, Department of Physics,
Budapest University of Technology and Economics, 1111-Budapest,
Hungary; e-mail: kezsmark@dept.phy.bme.hu.

\bibliography{malaria}

\end{document}